\documentclass[aps,twocolumn,
superscriptaddress,
footinbib,
prl]{revtex4-1}

\usepackage{amsmath,amssymb,bm,bbold}
\usepackage{mathtools}
\usepackage{graphicx}
\usepackage{multirow}
\usepackage{epstopdf}
\usepackage{latexsym}
\usepackage[normalem]{ulem} 
\usepackage[caption=false]{subfig}
\usepackage[usenames,dvipsnames]{color}
\usepackage{hyperref}
\usepackage{xcolor}
\usepackage{braket}
\usepackage{physics}
\usepackage{stackengine}
\usepackage[utf8]{inputenc} 
\usepackage{natbib}

\begin{document}

\title{Glassy quantum dynamics of disordered Ising spins}

\date{\today}

\author{P.~Schultzen}
\email{These authors contributed equally to this work. }
\author{T.~Franz}
\email{These authors contributed equally to this work. }
\author{S.~Geier}
\author{A.~Salzinger}
\author{A.~Tebben}
\author{C.~Hainaut}
\author{G.~Z\"urn}
\author{M.~Weidemüller}
\affiliation{Physikalisches Institut, Universit\"at Heidelberg, Im Neuenheimer Feld 226, 69120 Heidelberg, Germany}
\author{M.~G\"{a}rttner}
\email{Corresponding author. marting@kip.uni-heidelberg.de}
\affiliation{Physikalisches Institut, Universit\"at Heidelberg, Im Neuenheimer Feld 226, 69120 Heidelberg, Germany}
\affiliation{Kirchhoff-Institut f\"{u}r Physik, Universit\"{a}t Heidelberg, Im Neuenheimer Feld 227, 69120 Heidelberg, Germany}
\affiliation{Institut f\"ur Theoretische Physik, Ruprecht-Karls-Universit\"at Heidelberg, Philosophenweg 16, 69120 Heidelberg, Germany}

\begin{abstract}
We study the out-of-equilibrium dynamics in the quantum Ising model with power-law interactions and positional disorder. For arbitrary dimension $d$ and interaction range $\alpha \geq d$ we analytically find a stretched exponential decay of the global magnetization and ensemble-averaged single-spin purity with a stretch-power $\beta = d/\alpha$  in the thermodynamic limit. Numerically, we confirm that glassy behavior persists for finite system sizes and sufficiently strong disorder. We identify dephasing between disordered coherent pairs as the main mechanism leading to a relaxation of global magnetization, whereas genuine many-body interactions lead to a loss of single-spin purity which signifies the build-up of entanglement.
The emergence of glassy dynamics in the quantum Ising model extends prior findings in classical and open quantum systems, where the stretched exponential law is explained by a scale-invariant distribution of time scales, to both integrable and non-integrable quantum systems.
%We conclude that the magnetization decay is due to interaction induced dephasing while entanglement builds up at a smaller rate evident from the decay of single-spin purity, thus providing a microscopic understanding of glassy dynamics in a disordered closed quantum system.
\end{abstract}

\maketitle  

%\section{Introduction}
Statistical mechanics provides a well-established framework for describing the macroscopic properties of matter in thermal equilibrium. In contrast, no general theoretical framework exists for describing dynamics out of equilibrium. Of particular interest are extremely slow relaxation processes observed in disordered materials like spin glasses~\cite{Binder1986SpinQuestions, phillips_1996}.
Phenomenologically, the relaxation in these systems can be represented by a stretched exponential law $\exp[-(\gamma \tau)^\beta]$ with decay rate $\gamma$ and stretch-power $\beta$~\cite{1854AnP...167...56K}. Despite the widespread success of this heuristic description, a derivation of the stretched exponential law starting from first principles in a microscopic model has been derived for few systems only, in particular amorpheous solids~\cite{phillips_1996} and spin glasses~\cite{DeDominicis1985StretchedEnergies, phillips_1996}. By generalizing three prototypical models, Klafter and Shlesinger conjectured that a scale-invariant distribution of relaxation times is the unifying basis of stretched-exponential relaxation phenomena ~\cite{Klafter1986OnSystems}.

Recently, glassy dynamics has been found to emerge also in disordered quantum systems. Sub-exponential relaxation dynamics was observed in experiments with nitrogen-vacancy centers in diamond \cite{Choi2017DepolarizationEnsemble, Kucsko2018CriticalDiamond,davis2021probing} and in many-body localized systems under the influence of dissipation \cite{Everest2017RoleSystem}.
These studies involve \emph{open} quantum systems where dissipation arising from coupling to an external bath explains the slow relaxation. In the generic fluctuator model \cite{Choi2017DepolarizationEnsemble}, each particle is coupled to a local bath resulting in an average over different decay rates and thus a stretched-exponential decay law.
Recently, glassy relaxation was also found in a \emph{closed} quantum system governed by purely unitary dynamics \cite{Signoles2021GlassySystem}. In the absence of dissipation
%the system can only relax to a steady-state via interactions. Therefore
, the question arises whether and how glassy dynamics in isolated quantum systems is related to the degree of disorder and to the build-up of entanglement.

Understanding of the dynamics of strongly interacting disordered quantum many-body systems is notoriously difficult due to the lack of applicable theoretical approaches. The absence of a small parameter in the model impedes the use of perturbative methods, and the exponential complexity of quantum many-body problems generally limits numerical simulations to very small system sizes.
A paradigmatic exception is the quantum Ising model, where analytical solutions are available even for the disordered case \cite{Emch1966Non-markovianEquilibrium, Radin1970ApproachModel}. This model is diagonal in a product-state basis but, if prepared initially in a superposition of different eigenstates, features intrinsically quantum properties, namely dephasing between its eigenstate components leading to relaxation and the build-up of entanglement. 
Previous studies addressed the build-up of correlations \cite{Hazzard2014QuantumSystems}, decoherence \cite{Foss-Feig2013NonequilibriumSolution}, the effect of long-range interactions \cite{doi:10.1063/1.470956, Kastner2011DivergingModels, VanDenWorm2013RelaxationSimulatorb} and the decay of the Ramsey contrast \cite{Mukherjee2016AccessingDynamicsb, PhysRevA.94.053607}.

%Inspired by the conclusion of Klafter and Shlesinger~\cite{Klafter1986OnSystems}, we study the quantum Ising model with a scale-invariant distribution of interaction strengths. Here, we generalize previous approaches~\cite{doi:10.1063/1.470956,PhysRevA.94.053607} to analytically solve the dynamics of the transversal magnetization and purity. Results are provided for arbitrary dimensionality and power-law interaction, applicable in multiple experimental realizations, e.g. in  NMR \cite{PhysRevLett.67.2076} , quantum information \cite{Harris162,PhysRevLett.86.5188}, trapped ions  \cite{Britton2012EngineeredSpins}) and Rydberg atoms~\cite{Labuhn2016}. \\
%Remarkably, we find the same glassy dynamics with equal stretch power as the \textit{parallel channels model} from~\cite{Klafter1986OnSystems}. This model features the same spatial distribution and power-law interaction but instead of exponential relaxation, spins interact coherently.
%We identify dephasing between disordered coherent pairs as the main mechanism leading to a relaxation of global magnetization, whereas genuine many-body interactions lead to a loss of single-spin purity which signifies the build-up of entanglement.
%Furthermore, we find glassy dynamics also in the qunatum Ising model with a scale-invariant distribution of interactions hereby extending the work of Klafter and Shlesinger~\cite{Klafter1986OnSystems} to the quantum realm.

%Especially, we want to highlight previous work \cite{doi:10.1063/1.470956,PhysRevA.94.053607}, where stretched exponential behavior was shown to emerge in special cases.
Here, we introduce a generalized approach to obtain stretched exponential relaxation of the transversal magnetization and purity in the quantum Ising model extending earlier studies of special cases~\cite{doi:10.1063/1.470956,PhysRevA.94.053607}. Analytical results are provided for arbitrary dimensionality and power-law interactions, applicable to multiple experimental settings, e.g. in  NMR \cite{PhysRevLett.67.2076} , quantum information \cite{Harris162,PhysRevLett.86.5188}, trapped ions  \cite{Britton2012EngineeredSpins} and Rydberg atoms~\cite{Labuhn2016,Signoles2021GlassySystem}.
The analytic solution for the magnetization and purity of the microscopic model allows one to differentiate dephasing between disordered coherent pairs from genuine many-body effects. Furthermore, finding glassy dynamics in the quantum Ising model with a scale-invariant distribution of interactions constitutes a generalization of the Klafter-Shlesinger conjecture~\cite{Klafter1986OnSystems} to the quantum realm. 

%, where a stretched exponential behavior was found in special cases. Here we set these previous results in context to glassy dynamics and generalize them by analytically calculating the macroscopic relaxation dynamics of the magnetization for arbitrary dimensionality and power-law interactions. Moreover, we extend the method to investigate the growth of entanglement entropy which is expressed by the loss of ensemble-averaged single-spin purity. 

%\section{Ising spin dynamics with random positions}
%\label{Model}
We consider $N$ spin-1/2 particles, whose dynamics are governed by the Ising model ($\hbar=1$)
\begin{equation}
    \mathcal{H}_{\mathrm{Ising}} =  \sum_{i < k}J_{ik} \hat{\sigma}_z^i \otimes \hat{\sigma}_z^k  \,,
\end{equation}
where $\hat{\sigma}_\alpha^{i,k}$ ($\alpha = \{x,y,z\}$) are the Pauli operators acting on spin $i$ and $k$ and $J_{ik}$ describes the interaction between them. We consider isotropic power-law interactions $J_{ik} = C_\alpha/|\mathbf{r}_i - \mathbf{r}_k|^\alpha$ with particle positions $\mathbf{r}_i$, realized by a variety of quantum simulation platforms, such as polar molecules ($\alpha = 3$) \cite{Yan2013, Hazzard2014Many-bodyLattice}, Rydberg atoms  ($\alpha=3,6$) \cite{Orioli2018, Signoles2021GlassySystem, Scholl2020, Ebadi2020}  or trapped ions ($0\leq \alpha < 3$) \cite{Britton2012EngineeredSpins, Monroe2019} (see \cite{Hazzard2014QuantumSystems} for a more complete list).
For the initial state  $\ket{\psi_0} = \ket{\rightarrow} ^{\otimes N}$, we are interested in the relaxation of the ensemble-averaged transversal magnetization $\overline{\langle \hat{s}_x \rangle} = N^{-1} \sum_i \langle \hat{\sigma}_x^i \rangle/2$, where the overline denote the ensemble average and $\Braket{\dots}$ the quantum mechanical expectation value. Here, $\ket{\rightarrow}$ is the $\hat{\sigma}_x$ eigenstate with $\hat{\sigma}_x \ket{\rightarrow} = \ket{\rightarrow} $.

The dynamics of Ising spins initialized in $\ket{\psi_0}$ was described analytically by Emch \cite{Emch1966Non-markovianEquilibrium} and Radin \cite{Radin1970ApproachModel} as
\begin{equation}
\label{EmchRadin}
    \overline{\langle \hat{s}_x(\tau) \rangle} = \frac{1}{2} \sum_i \frac{1}{N}\langle \hat{\sigma}^i_x(\tau)\rangle  =\frac{1}{2} \sum_i \frac{1}{N}  \prod_{k \neq i}^N \cos(2 J_{ik} \tau) \,,
\end{equation}
which shows that the ensemble average is determined by products of oscillations with frequencies given by the couplings $J_{ik}$ between a given spin $i$ and its neighbors $k$. From the Emch-Radin solutions also follows $\langle \hat{\sigma}_y^i \rangle = \langle \hat{\sigma}_z^i \rangle = 0$ such that the analytical expression  $\langle \hat{\sigma}^i_x(\tau)\rangle  = \prod_{k \neq i}^N \cos(2 J_{ik} \tau)$ already fully determines the one-particle reduced density matrix $\rho^i= \left( \mathbb{1} + \langle \hat{\sigma}_x^i \rangle \hat{\sigma}_x^i \right)/2$ of spin $i$.
Thus, the single-particle purity is 
\begin{equation}
    \mathrm{tr}[(\rho^i)^2] = \frac{1}{2}\left(1 + \langle \hat{\sigma}_x^i(\tau) \rangle^2 \right) \,.
\end{equation}
Similar to the magnetization, we define the ensemble-averaged single-particle purity as
$\overline{\mathrm{tr}(\rho^{2})} = \frac{1}{2}\left(1 + \overline{\langle \hat{\sigma}_x(\tau) \rangle^2} \right)$, where 
\begin{equation}
\label{purity}
    \overline{\langle \hat{\sigma}_x(\tau) \rangle^2} = \sum_i \frac{1}{N}\langle \hat{\sigma}_x^i(\tau) \rangle^2  = \sum_i \frac{1}{N}\prod_{k \neq i}^N \cos^2(2 J_{ik} \tau) \,.
\end{equation}
The purity of a subsystem (here, a single spin) of a closed quantum system $1/2 \leq  \mathrm{tr}[(\rho^i)^2] \leq 1 $ quantifies the entanglement between the subsystem and its complement, and determines the second order Rényi entropy $S_2=-\log( \mathrm{tr}[(\rho^i)^2])$. For our initial product state $S_2=0$, as the single-particle reduced state $\rho^i$ is pure, and the subsystem entropy takes its maximal value $S_2=\log(2)$ in the late-time limit.

\begin{figure}
\includegraphics[width= \linewidth]{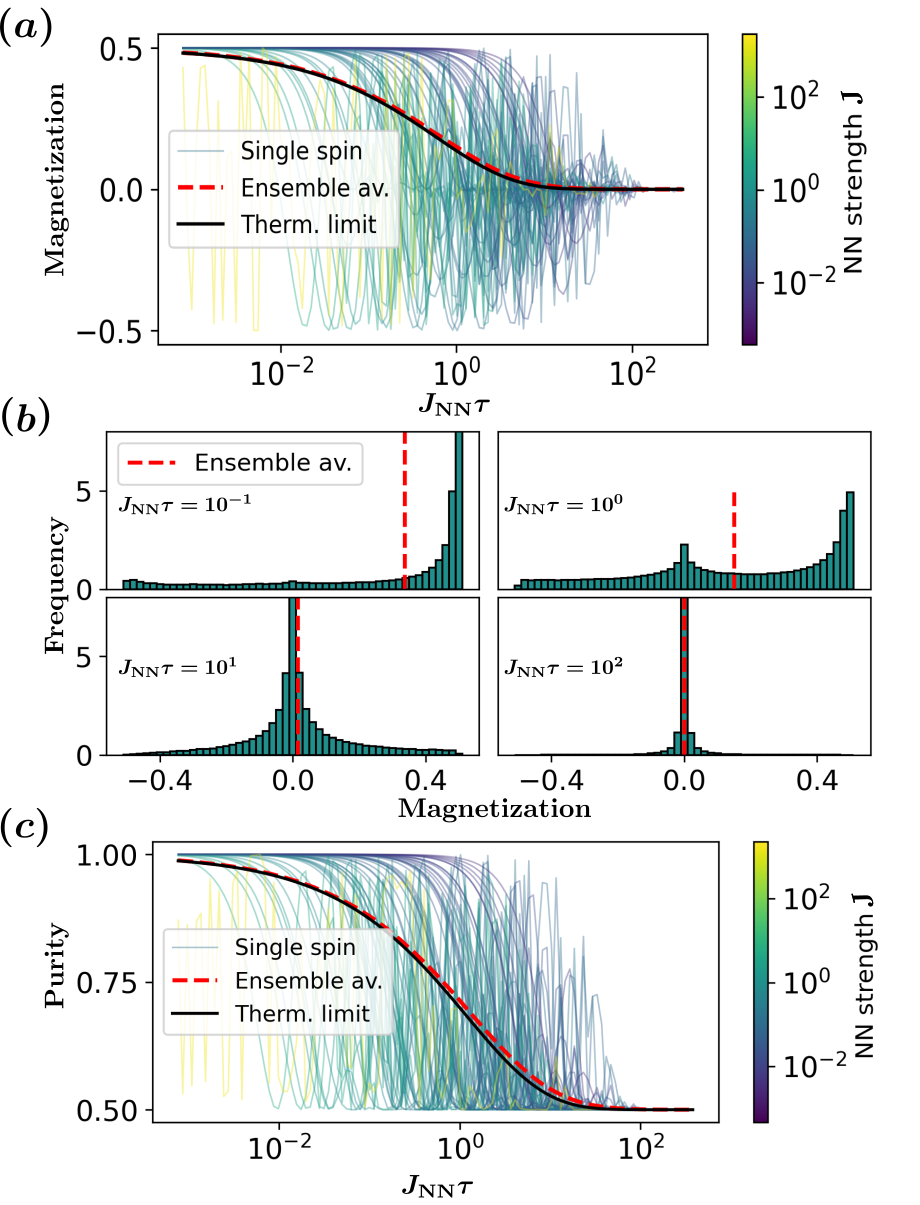}
\caption{$\bm{(a)}$ Magnetization decay for a uniform random spin distribution in $d=3$ with $\alpha = 6$ for 1300 spins. The single-spin magnetizations $\langle\hat s_x^i(\tau)\rangle$ for $50$ different spins are shown, featuring the oscillatory behavior predicted by Eq.~\eqref{EmchRadin}. The line color encodes the nearest-neighbor (NN) interaction strength and is thus a measure for the onset and fastest frequency of each oscillation. Its median $J_{\mathrm{NN}}$ is furthermore used as the unit for the relaxation time. Additionally, the ensemble-averaged decay is plotted (red dashed line), showing monotonous sub-exponential relaxation, which is well captured by a stretched exponential function predicted analytically in the large $N$ limit (black line). Remaining deviations from the analytical solution can be attributed to the finite system size used for the simulation. $\bm{(b)}$ Histograms showing the frequency of occurrence of single-spin magnetization values at different times. The fluctuations relax on a slower time scale than the mean value, which is directly connected to the decay of the purity. $\bm{(c)}$ Single-spin purities and ensemble average, analogous to $\bm{(a)}$.}
\label{figure1}
\end{figure}

The Emch-Radin solutions hold for arbitrary choices of the couplings $J_{ik}$. 
Here, we consider disorder in the couplings due to random spin positions drawn from a uniform distribution within a $d$-dimensional sphere and power-law interaction with exponent $\alpha \geq d$. To illustrate the characteristic dynamics emerging in this situation we show the relaxation of the transversal magnetization in Fig.~\ref{figure1}(a) for  Van der Waals interaction ($\alpha$ = 6) in $d=3$ dimensions. Time is scaled by the median nearest-neighbour (NN) interaction strength $J_{\mathrm{NN}}$ \cite{Signoles2021GlassySystem}. The random positions lead to a strongly disordered $J_{ik}$ distribution which causes oscillations on a broad range of different time scales. Curves showing fast oscillations correspond to spins interacting strongly with their nearest neighbors. Due to disorder, these oscillations between coherent pairs loose their phase correlations. Consequently, the ensemble-averaged magnetization (red dashed line) shows smooth sub-exponential decay closely following the analytical solution in the thermodynamic limit $N \rightarrow \infty$ (black curve) derived below, which is a stretched exponential function.

Panel~(b) shows the frequency of occurrence of single-spin magnetizations at fixed evolution times, showing a bimodal distribution at intermediate times. For $J_{\mathrm{NN}}\tau  = 10$, the ensemble-averaged value nearly reached its equilibrium, while still showing large fluctuations around the mean value. These fluctuations are directly connected to the purity (cf.\ Eq.~\eqref{purity}) showing already that the decay of magnetization and purity happen on different time scales. In this particular case, the time scales differ by a factor of 2, where in general this factor depends on $\alpha/d$ and can become large as discussed below. Panel~(c) shows the ensemble-averaged purity along with the purity of individual spins. Similar to panel (a), the purity of individual spins shows oscillations. Following the same argument that explains the full relaxation of magnetization, the dephasing of these oscillations would result in an average purity of 0.75. Instead, the ensemble averaged purity relaxes to its minimum value of $0.5$, which accounts for an irreversible build-up of entanglement with the whole ensemble. We again find a smooth stretched exponential curve in the ensemble average.

%\section{Analytical results for ensemble-averaged quantities}
To derive an analytical expression for Eq.~\eqref{EmchRadin} in the limit of $N\to \infty$, the ensemble average can be replaced by an average over all possible configurations of placing the surrounding spins of a reference spin \cite{PhysRevA.94.053607, Mukherjee2016AccessingDynamicsb} thus leading to a scale-invariant distribution of interaction strengths. Without loss of generality, we fix the position of the reference spin at $\mathbf{r}_1=0$ and choose a finite spherical integration volume $V$ in which $N^\prime$ atoms are placed. We will later take the limit $N^\prime \to\infty$ keeping the density $n=N^\prime/V$ constant. 
Therefore, Eq.~\eqref{EmchRadin} transforms into the integral form
\begin{equation}
\label{eq:cont_general}
    \overline{\langle \hat{s}_x(\tau) \rangle}  = \frac{1}{2} \int_V  d\mathbf{r}_2\ldots d\mathbf{r}_{N^\prime} P( \mathbf{r}_2\ldots \mathbf{r}_{N^\prime}) \prod_{k=2}^{N^\prime} \cos(2 J_{1k} \tau) \,.
\end{equation}
The spin positions are chosen independently following a homogeneous distribution over the volume $V$, i.e.\ $P( \mathbf{r}_2\ldots \mathbf{r}_{N^\prime}) = \prod_k p(\mathbf{r}_k)$ with $p(\mathbf{r}_k)=1/V$. Thus, the integral in Eq.~\eqref{eq:cont_general} factorizes into a product of identical integrals
\begin{equation}
\label{eq:cont_factored}
    \overline{\langle \hat{s}_x(\tau) \rangle}  = \frac{1}{2} \left[\frac{1}{V}\int_V  d\mathbf{r} \cos(2 J_{\mathbf{r}} \tau)\right]^{N^\prime-1} \,,
\end{equation}
where $J_{\mathbf{r}} = C_\alpha/|\mathbf{r}|^\alpha$.

We now introduce a lower distance cutoff $r_b$ on the integration volume \footnote{This is motivated physically: Arbitrarily closely spaced spins would have arbitrarily large interaction strength requiring a high energy cutoff. Also, in experimental realizations with Rydberg atoms, a natural lower distance cutoff is given by the dipole blockade radius \cite{Comparat2010}.}. Note that imposing an exclusion distance $r_b$ between any pair of atoms violates the assumption of independent atom positions and scale-invariant distributions of interactions. 
For our analytical calculations this inconsistency is irrelevant as we will send $r_b$ to zero eventually. We show below that our results also describe the dynamics well for finite exclusion radius as long as $r_b$ is much smaller than the average nearest neighbor distance in the ensemble.

Defining $r_0$ as the radius of the spherical integration volume $V$ and carrying out the angular part of the integration we obtain
\begin{equation}
\label{eq:int_start}
    \overline{\langle \hat{s}_x(\tau) \rangle}  = \frac{1}{2}\left[\frac{d}{r_0^d - r_b^d} \int_{r_b}^{r_0} dr\, r^{d-1} \cos\left(2\frac{ C_\alpha}{r^{\alpha}} \tau\right)\right]^{N'-1} \,.
\end{equation}
We now evaluate this expression in the limits $r_b\to 0$ and $r_0, N^\prime\to\infty$ for arbitrary $d$ and $\alpha\geq d$, thus generalizing previous results. The scale invariance of the system now becomes obvious as Eq.~\eqref{eq:int_start} is invariant under a rescaling of space ($r\rightarrow \lambda r$) and time ($\tau\rightarrow \lambda^\alpha \tau$).
The main result of our derivation
\begin{equation}
\label{eq:Sx_result}
    \overline{\langle \hat{s}_x(\tau) \rangle} = \frac{1}{2} \exp\left[-\kappa_{d,\alpha} \Gamma\left(\frac{\alpha -d}{\alpha}\right) \sin (\pi \frac{\alpha -d }{2\alpha })\tau^{d/\alpha} \right]
\end{equation}
is a stretched exponential $\overline{\langle \hat{s}_x(\tau) \rangle} = \exp\left[-(\gamma_m \tau )^{\beta_m}\right]/2$
with decay rate $\gamma_m  = \left[\kappa_{d,\alpha} \Gamma\left(\frac{\alpha -d}{\alpha}\right) \sin (\pi \frac{\alpha -d }{2\alpha }) \right]^{\alpha/d}$
and stretch power $\beta_m = \frac{d}{\alpha}$ (for details see Supplemental Material (SM) \cite{SM} containing references~\cite{mathar2012series}). Here, the index $m$ stands for magnetization and we have introduced $\kappa_{d,\alpha} = \pi ^{d/2} n  (2 C_\alpha)^{d/\alpha}/\Gamma(d/2 +1)  $.
Since $\beta_m\leq 1$ our result shows that the characteristic sub-exponential relaxation typically observed in glassy systems appears in the out-of-equilibrium unitary dynamics under the Ising Hamiltonian.
In the case $\alpha = d$ Eq.~\eqref{eq:Sx_result} simplifies to a pure exponential decay $\overline{\langle \hat{s}_x(\tau) \rangle}_{\alpha = d} = \exp\left(- \pi \kappa_{d,\alpha}  \tau/2 \right)/2$
where we used, that $\lim_{\alpha -d  \to 0}  \left[\Gamma\left(\frac{\alpha -d}{\alpha}\right) \sin \left(\pi \frac{\alpha -d }{2\alpha }\right)\right] = \pi / 2$. Note, that the derivation of a stretched exponential remains valid even for a broad class of anisotropic interactions, whose anisotropy yields a change only in the rate $\gamma$, whereas $\beta$ remains unchanged (see SM \cite{SM}).

Remarkably, the stretch power $\beta = d/\alpha$ is the same as for the \textit{Förster Direct-Transfer Model} with parallel channels discussed by Klafter and Shlesinger~\cite{Klafter1986OnSystems}. This classical model features the same spatial distribution and power-law interaction but relies on exponential relaxation, instead of coherently interacting spins showing microscopic oscillatory behaviour. 

Beyond classical models, genuine quantum effects occur in the quantum Ising model. Therefore, we focus on the ensemble-averaged purity which describes the build-up of entanglement, cf.\ Eq.~\eqref{purity}. As for the magnetization, one can convert the ensemble average of the term $\overline{\langle \hat{\sigma}_x(\tau) \rangle^2}$ into an integral over atom positions in the asymptotic large $N$ limit, resulting in
\begin{equation}
\begin{split}
        \overline{\langle \hat{\sigma}_x(\tau) \rangle^2} &= \left[\frac{d}{r_0^d - r_b^d} \int_{r_b}^{r_0} \! dr \, r^{d-1} \cos^2\left(2 \frac{ C_\alpha}{r^{\alpha}} \tau\right) \right]^{N^\prime-1} \\
        &= \left[\frac{1}{2} + \frac{d}{r_0^d - r_b^d} \int_{r_b}^{r_0} \! dr \frac{r^{d-1}}{2} \cos\left(4 \frac{ C_\alpha}{r^{\alpha}} \tau\right) \right]^{N^\prime-1} \,,
\end{split}
\end{equation}
where the identity $\cos^2(x) = 1/2 + \cos(2x)/2$ is used. The integral now has the same shape as the one for the magnetization \eqref{eq:int_start} with a global prefactor of $1/2$ and twice the frequency $C_\alpha \rightarrow 2 C_\alpha$. We can thus use the same approach to obtain 
\begin{equation}
    \overline{\mathrm{tr}(\rho^{2})} = \frac{1}{2}\left\{1+ \exp\left[-(\gamma_p \tau)^{\beta_p}\right]\right\}
\end{equation}
for the relaxation of the ensemble-averaged single-particle purity with $\gamma_p = 2^{1-\alpha/d}  \gamma_m$ and $\beta_p = \beta_m = d/\alpha$. This formalism can be extended to all higher moments $\overline{\langle \hat{\sigma}_x(\tau) \rangle^j}$ with $j \in \mathbb{N}$ shown in SM~\cite{SM}. Note that the decay rate of the purity is generally smaller than that of the magnetization by a factor $\gamma_p/\gamma_m= 2^{1-\alpha/d}\leq 1$.
The slower decay of purity is visible in the fluctuations of the single-spin magnetizations (cf. Fig. \ref{figure1}(b)) that are still present when the mean magnetization has already decayed. This separation of time scales gets large in the case of $\alpha \gg d$.

We numerically investigate, whether glassy dynamics persist for systems with finite exclusion radius $r_b$ and finite system size $N$. We evaluate Eqs.~\eqref{EmchRadin} and \eqref{purity} for $d=1,2,3$, $\alpha = d,\dots,10$ and random atom positions. We average the results over $N_s$ random realizations to decrease statistical fluctuations from random sampling and fit the averaged relaxation curves with a general stretched exponential function described by $f(\tau) = A \exp\left[ - (\gamma \tau)^\beta\right]$
and we compare the resulting $\beta$ to the analytical solution $d/\alpha$ derived previously of the thermodynamic limit. 

The exclusion radius $r_b$ is incorporated in the process of generating random position samples by rejecting atoms that are closer than $r_b$ to one of their neighbors. This process is equivalent to the random sequential absorption (RSA) model of randomly placing non-overlapping spheres \cite{Adamczyk1996, Hinrichsen1990}. The packing density can be quantified by the ratio $x = N r_b^d/r_0^d$, where a small value of $x$ corresponds to strong disorder, i.e. uncorrelated atom positions, while large $x$ implies more densely packed and thus more regularly spaced, less disordered spins. 
We note that in experiments with Rydberg atoms $x$ is tunable over a wide range \cite{Signoles2021GlassySystem}.

\begin{figure}[h!]
\includegraphics[width= \linewidth]{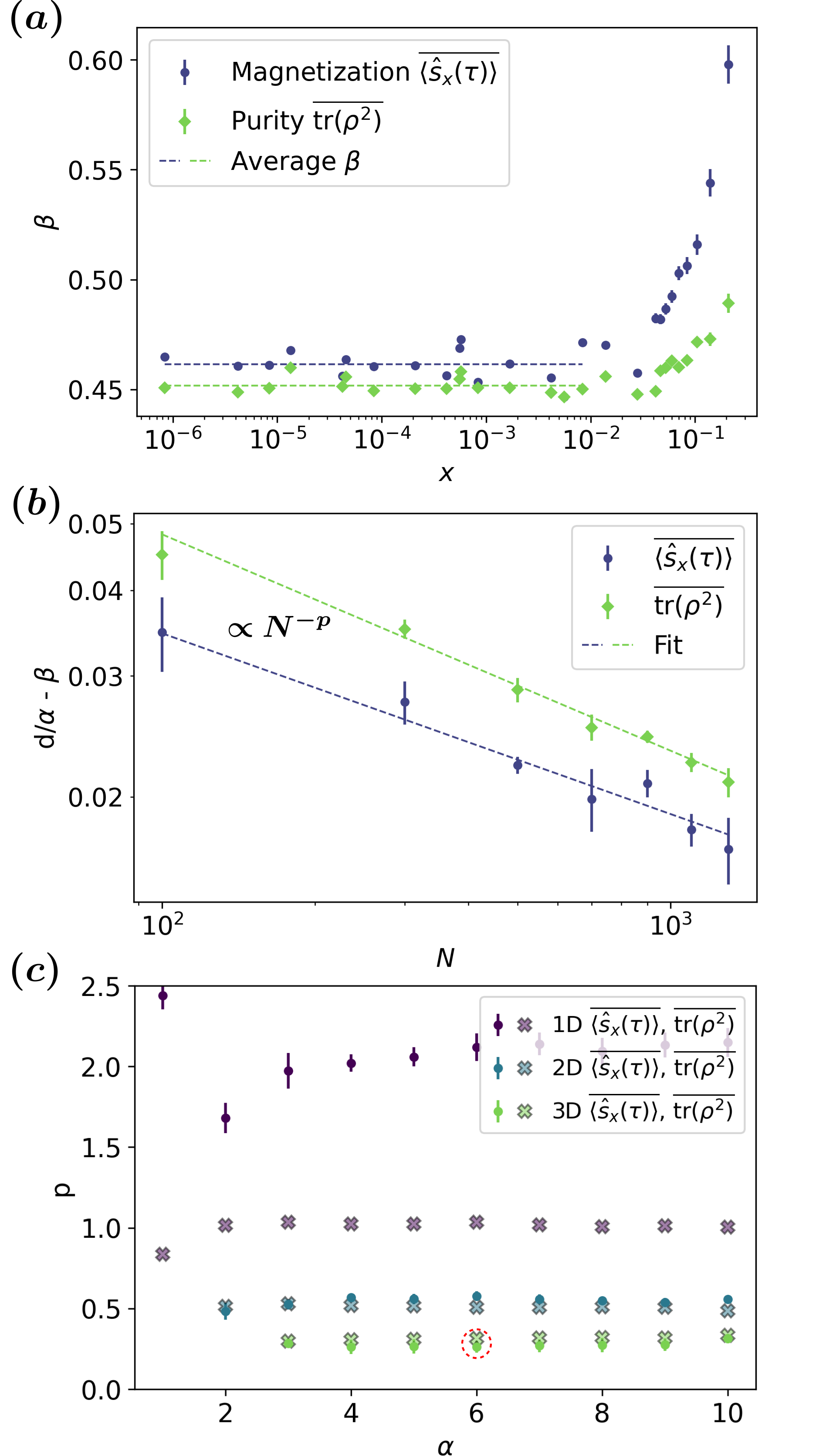}
\caption{$\bm{(a)}$ Fit parameter $\beta$ as a function of the disorder parameter $x = N r_b^d/r_0^d$ for the magnetization (blue dots) and purity (green dots) for $N = 100$ and $N_s=200$ for the case of $\alpha = 6$ in $d=3$ . At strong disorder ($x \lesssim 0.01$) $\beta$ becomes independent of $x$. Dashed lines show the averages within this regime. Error bars describe the parameter uncertainty of $\beta$ obtained from the fit. $\bm{(b)}$ Deviation of the fitted $\beta$ from the analytical solution $d/\alpha = 0.5$ as a function of $N$. We average $\beta$ over five different blockade radii within the strongly disordered regime. The error bars correspond to the standard error of the mean of the latter, which is the dominant uncertainty, cf.~($\bm{a}$). Dashed lines are power law fits. $\bm{(c)}$ Fitted power law exponent $p$ for all simulated cases $d=1,2,3$ and $\alpha = d,\dots,10$. The red circle highlights the points corresponding to the data shown in $\bm{(b)}$. See Supplemental Material for the choice of parameters $N$ and $N_s$. 
}
\label{figure2}
\end{figure}

The dependence of $\beta$ on $x$ is shown in Fig.~\ref{figure2}(a) for both magnetization and purity in the case of Van der Waals interaction $\alpha = 6$ and $d = 3$ for a system size of $N=100$ and $N_s=200$ samples. In the sufficiently disordered regime ($x \lesssim 0.01$) $\beta$ reaches a constant value (dashed lines), which shows, that the description by glassy dynamics obtained in the limit $r_b \to 0$ are robust with respect to finite exclusion radius. In this regime the blockade radius is sufficiently small, such that the system can be considered as effectively scale invariant. Similar results are obtained in all studied cases of dimension and interaction range.

Next we study the effect of finite $N$ in the strongly disordered regime ($x\ll 1$). Fig.~\ref{figure2}(b) shows the deviation of the fitted $\beta$ from the analytical result $d/\alpha$ as a function of $N$. Analogous plots for $\alpha=6$ in $d=1$ and $d=2$ dimensions are shown in the SM \cite{SM}. We observe an algebraic decrease of the error for both magnetization and purity. A power-law fit $\propto N^{-p}$ shows good agreement. The point at $N=1300$ corresponds to the data shown in Figs.~\ref{figure1}(a) and (c), where the comparison to the analytical solution matches nearly perfectly. 

We systematically extract the exponent $p$ describing the scaling of the error with $N$ for various $d$ and $\alpha$ (see SM \cite{SM} for the choice of parameters $N$ and $N_s$). The range of particle numbers is chosen such that the deviation from the analytical solution does not fall below $\sim 1\%$. This value corresponds to the size of statistical fluctuations due to finite disorder averaging giving a lower bound on the observable deviation. In particular the 1D case converges already for small $N$, therefore we need to increase the samples $N_s$ to reduce statistical fluctuations. The results, shown in Fig.~\ref{figure2}(c) indicate that the finite size scaling behavior is independent of $\alpha$, but convergence is slower for increasing $d$. In all cases, an algebraic convergence to the analytical result is obtained, showing the robustness of our analytical results with respect to finite size effects. 

%\section{Conclusions}
Our analytical and numerical studies show that the far-from-equilibrium dynamics of the quantum Ising model exhibits glassy behavior. Additional to the global magnetization, we investigated the single-spin purity, which quantifies entanglement between local spins and their environment. 
Especially for short interaction range $\alpha \gg d$, the time scales between magnetization and purity differ largely. This discrepancy is due to differences in the involved relaxation mechanisms. For qualitatively explaining the decay of the global magnetization it is sufficient to consider the interaction of spins with their nearest neighbors. Due to disorder, these coherent pair dynamics oscillate at different frequencies resulting in dephasing and hence a loss of global magnetization. 
In contrast, the full relaxation of single-spin purity is a genuine many-body effect, which is eventually due to the irreversible dephasing between many-body eigenstates.

Similar to known classical models showing glassy dynamics, the quantum Ising model features a scale-invariant distribution of time scales. Therefore, our findings extend the conclusion of \cite{Klafter1986OnSystems}, that scale-invariance is sufficient to explain the emergence of a stretched exponential law, to the quantum realm. This argument is not limited to observables like the magnetization with a classical analogue, but also applies to the genuine quantum effect of purity which shows the same stretched exponential relaxation. An interesting direction for future research is the investigation of the dynamics of entanglement entropy beyond single-spin subsystems. This includes entanglement scaling with subsystem size \cite{Abanin2019MBLEntanglement, Hertzberg2011Wilcek} in view of constraints on the spreading of correlations \cite{Lieb1972TheSystems,PhysRevLett.113.030602,PhysRevLett.114.157201,Hastings2006}.

%Regarding the conclusion of \cite{Klafter1986OnSystems}, that scale-invariant relaxation times are sufficient for the observation of stretched exponential behavior, our analytical derivation of glassy dynamics extends the known classical models to the quantum Ising model, where the scale-invariance is found in the interaction distribution. The applicability to quantum properties is further strengthened by the observation of sub-exponential decay of the purity, which is a pure quantum effect without a classical analogue.

In conclusion, stretched exponential relaxation is found in classical models as well as in open quantum systems and, as we have shown, also in the quantum Ising model prototypical for isolated integrable quantum systems. Despite the vastly different underlying physics, all of these systems feature scale-invariant distributions of time scales. %In addition to the result of stretched exponential decay in the quantum Ising model, previous works focusing on open quantum systems explained its occurrence by the coupling to a locally varying Markovian bath, which also results in scale-invariant dynamics. \\
Thus, the analytical results presented here are in line with the conclusion of \cite{Klafter1986OnSystems}, thus extending the sufficiency of scale-invariance for the emergence of glassy dynamics to quantum systems. Based on recent numerical investigations of a more general family of Heisenberg Hamiltonians, where glassy dynamics is observed for almost any anisotropy parameter \cite{schultzen2021semiclassical}, as well as experimental findings~\cite{Signoles2021GlassySystem}, we expect the conjecture to hold even for non-integrable quantum systems if scale-invariance is given.
We thank Adrian Braemer and Peter Kaposvari for helpful discussions.
This work is supported by the Deutsche Forschungsgemeinschaft (DFG, German Research Foundation) under Germany’s Excellence Strategy EXC2181/1-390900948 (the Heidelberg STRUCTURES Excellence Cluster), within the Collaborative Research Center SFB1225 (ISOQUANT) and the DFG Priority Program 1929 “GiRyd” (DFG WE2661/12-1). 
We acknowledge support by the European Commission FET flagship project PASQuanS (Grant No. 817482) and by the Heidelberg Center for Quantum Dynamics. C.H. acknowledges funding from the Alexander von Humboldt foundation and T.F. from a graduate scholarship of the Heidelberg University (LGFG).

\bibliography{references, refs_martin}

\section{Supplementary materials}
\section{Analytical derivation}
\label{supA}
In order to shorten notation we abbreviate the oscillation frequency of the cosine term in Eq.~(7) of the main text as $\omega  \coloneqq 2 C_\alpha/r^\alpha$ and let $\omega_0$ and $\omega_b$ be its value
at $r=r_0$ and $r=r_b$, respectively. In addition, we introduce the dimensionless variables $y_{0,b} \coloneqq (\omega_{0,b} \tau)^{1/\alpha}$. In terms of these new variables, the limits we are interested in are $\omega_b,y_b\to \infty$ and $\omega_0,y_0\to 0$.
Substitution of integration variables, integration by parts, and splitting up the integration interval $\int_{y_0}^{y_b}=\int_{0}^{y_b}-\int_{0}^{y_0}$ brings Eq.~(7) into the form
\begin{equation}
\label{eq:Sx_part_int}
\begin{split}
             & \overline{\langle \hat{s}_x(\tau) \rangle} =  \frac{1}{2}\left[\left( \frac{\omega_b \omega_0}{\omega_b - \omega _0}\right) ^{d/\alpha }\left( - \frac{\cos(\omega \tau)}{\omega^{d/\alpha}}  \Big|_{\omega_0}^{\omega_b} \right.\right.  \\
            & \left.\left. - \alpha \tau ^{d/\alpha}\left( \frac{\omega_b \omega_0}{\omega_b - \omega _0}\right) ^{d/\alpha }\left(I_{\mu\nu}(y_b) - I_{\mu\nu}(y_0)\right)\right)\right]^{N'-1},
\end{split}
\end{equation}
where we defined the integrals $I_{\mu\nu}(x) =  \int_{0}^{x} dy \, \sin(y^\mu) y^{\nu}$, where $\mu = \alpha$ and $\nu = \alpha - d -1$ in Eq.~\eqref{eq:Sx_part_int}. 

Next, we will take the limit $\omega_b \to \infty$, corresponding to $r_b\to 0$. For this, we 
Taylor expand the terms inside the square brackets in Eq.~\eqref{eq:Sx_part_int} at small $(\omega_0/\omega_b)^{d/\alpha}$. To first order we obtain
\begin{equation}
\label{Taylor1}
    -\left( \frac{\omega_b \omega_0}{\omega_b - \omega _0}\right) ^{d/\alpha }\frac{\cos(\omega \tau)}{\omega^{d/\alpha}}  \Big|_{\omega_0}^{\omega_b} \approx 1 + \frac{1 - \cos(\omega_b \tau)}{\omega_b^{d/\alpha}}\frac{\kappa_{d,\alpha}}{N'}
\end{equation}
and
\begin{equation}
\label{Taylor2}
    \left( \frac{\omega_b \omega_0}{\omega_b - \omega _0}\right)^{d/\alpha}  \left[I_{\mu\nu}(y_b) - I_{\mu\nu}(y_0)\right] 
     \approx \frac{\kappa_{d,\alpha}}{N'}I_{\mu\nu}(y_b) \,,
\end{equation}
where we have defined the constant $\kappa_{d,\alpha}$ via
\begin{equation}
\label{VolumeInequ}
    \omega_0^{d/\alpha} = \frac{\pi ^{d/2} n  (2 C_\alpha)^{d/\alpha}}{\Gamma(d/2 +1) N'}  \eqqcolon \frac{\kappa_{d,\alpha}}{N'}
\end{equation}
and the number density $n=N^\prime/V$, where $V$ is the volume of a $d$-dimensional sphere with radius $r_0$. With this definition we can later take the limit $r_0\to \infty$ by taking $N^\prime\to\infty$ and treating $\kappa_{d,\alpha}$ as a constant which ensures that $n$ is kept constant as required. 
We note that the Taylor series of $I_{\mu\nu}(y_0)$ is only convergent for $\alpha \geq d$.
We can now take the limit $\omega_b \to \infty$. The second term of Eq.~\eqref{Taylor1} vanishes. The integral in Eq.~\eqref{Taylor2} becomes
\begin{equation}
    \lim_{y_b\to \infty} I_{\mu\nu}(y_b) = \frac{1}{\alpha} \Gamma\left(\frac{\alpha -d}{\alpha}\right) \sin( \pi \frac{\alpha -d }{2\alpha }) \,,
\end{equation}
where the asymptotes of the generalized Fresnel integrals \cite{mathar2012series} are used. Assembling everything, the relaxation can be written as 
\begin{equation}
\label{penultimate}
     \overline{\langle \hat{s}_x(\tau) \rangle} =  \frac{1}{2}\Big[1 - \tau^{d/\alpha}  \frac{\kappa_{d,\alpha}}{N'} \Gamma\left(\frac{\alpha -d}{\alpha}\right) \sin (\pi \frac{\alpha -d }{2\alpha })\Big]^{N'-1} \,. 
\end{equation}

Finally, we take the limit $N' \to \infty$, corresponding to the infinite volume limit $r_0\to\infty$ which yields
\begin{equation}
\label{eq:Sx_result_supp}
    \overline{\langle \hat{s}_x(\tau) \rangle} = \frac{1}{2} \exp\left[-\kappa_{d,\alpha} \Gamma\left(\frac{\alpha -d}{\alpha}\right) \sin (\pi \frac{\alpha -d }{2\alpha })\tau^{d/\alpha} \right]. 
\end{equation}
\section{Generalization to anisotropic interaction}
So far, the interaction was restricted to isotropic interactions. However, the approach can be easily extended to anisotropic interactions as long as the interaction $J(\mathbf{r})$ can be factorized into an angular and radial component $J(\mathbf{r}) = f(\Omega) J(|\mathbf{r}|)$ where $\Omega$ is the $d$-dimensional solid angle and we keep a power law dependence on the distance $r = |\mathbf{r}|$ as before. Similar to Eq.~(6) of the main text, we obtain in spherical coordinates for anisotropic interaction
\begin{equation}
\begin{split}
            \overline{\langle \hat{\sigma}_x(\tau) \rangle}  = & \\
            \bigg[ \frac{d}{r_0^d - r_b^d} & \frac{1}{V_\Omega} \int_\Omega d\Omega \, \int_{r_b}^{r_0} dr\, r^{d-1} \cos\left(2 f(\Omega) \frac{ C_\alpha}{r^{\alpha}} \tau\right)\bigg]^{N'-1} \,,
\end{split}
\end{equation}
where we introduce the angular volume $V_\Omega = \int_\Omega d\Omega$. Note, that this expression is independent of the sign of $f(\Omega)$, which will be thus replaced by $|f(\Omega)|$. This simplifies further transformation and the following steps can be done similarly to the isotropic case. The variable transformation changes to $\omega_\Omega \coloneqq 2 |f(\Omega)| C_\alpha/r^\alpha$ and $y_\Omega$ accordingly.\\
As $|f(\Omega)|$ transforms equivalently to $\tau$, we obtain 
a stretched exponential function with new timescale $\gamma'$ given by 
\begin{equation}
    \gamma'=\chi \gamma
\end{equation}
where we define the impact of the anisotropy $\chi$ as 
\begin{equation}
    \chi \coloneqq \frac{1}{V_\Omega}  \int_\Omega d\Omega \, |f(\Omega)|^{d/\alpha} \, .
\end{equation}
Here $\gamma$ is the timescale of the similar isotropic  case and the resulting $\beta$ remains the same. \\
As an example we investigate the case of dipolar interaction in $d=3$ dimensions
\begin{equation}
    \label{dipolar}
    J(\mathbf{r})_{\mathrm{Dipolar}} = (1-3\cos^2(\Theta)) \frac{C_3}{|\mathbf{r}|^3} \, ,
\end{equation}
where we can identify $f(\Omega) = 1-3\cos^2(\Theta)$, which only depends on the azimuthal angle $\Theta$. We obtain $\chi = 4/(3\sqrt{3}) \sim 0.77$. \\

\section{Generalization to higher moments}
Using the formalism derived in the previous section, we can generalize the results to arbitrary moments $\overline{\langle \hat{\sigma}_x(\tau) \rangle^j}$ with $j \in \mathbb{N} $. For a single particle $i$ we obtain 
\begin{equation}
        \langle\hat{\sigma}_x^i(\tau)\rangle^j = \prod_{k\neq i}\cos^j(2J_{ik}\tau)
\end{equation}
and therefore by averaging over all configurations (cf. Eq.~(6) of the main text)
\begin{equation}
    \overline{\langle \hat{\sigma}_x(\tau) \rangle^j} = \left[\frac{d}{r_0^d - r_b^d} \int_{r_b}^{r_0} dr\, r^{d-1} \cos^j\left(2\frac{ C_\alpha}{r^{\alpha}} \tau\right)\right]^{N'-1} \,.
\end{equation}
Using the relation for the cosine 
\begin{equation}
    \cos(x)^j = \sum_{i=0}^j \binom{j}{i}\frac{\cos{((j-2i)x)}}{2^j}
\end{equation}
the integral is reduced to integrals over cosine terms, which were already evaluated in the previous section of the supplement. The procedure can be followed until Eq.~(6), where the different frequencies lead to an effective timescale $\gamma_j$ defined by 
\begin{equation}
    \gamma_j= \sum_{i=0}^j \left[ \binom{j}{i}^{\alpha/d }\Big(\frac{1}{2^j}\Big)^{\alpha/d}     |j-2i| \right]\gamma_m
\end{equation}
with the definition of $\gamma_m$ of the main text. 

\section{Additional numerical data and parameters}
\label{supB}

\label{supC}
\begin{figure}
    \includegraphics[width= \linewidth]{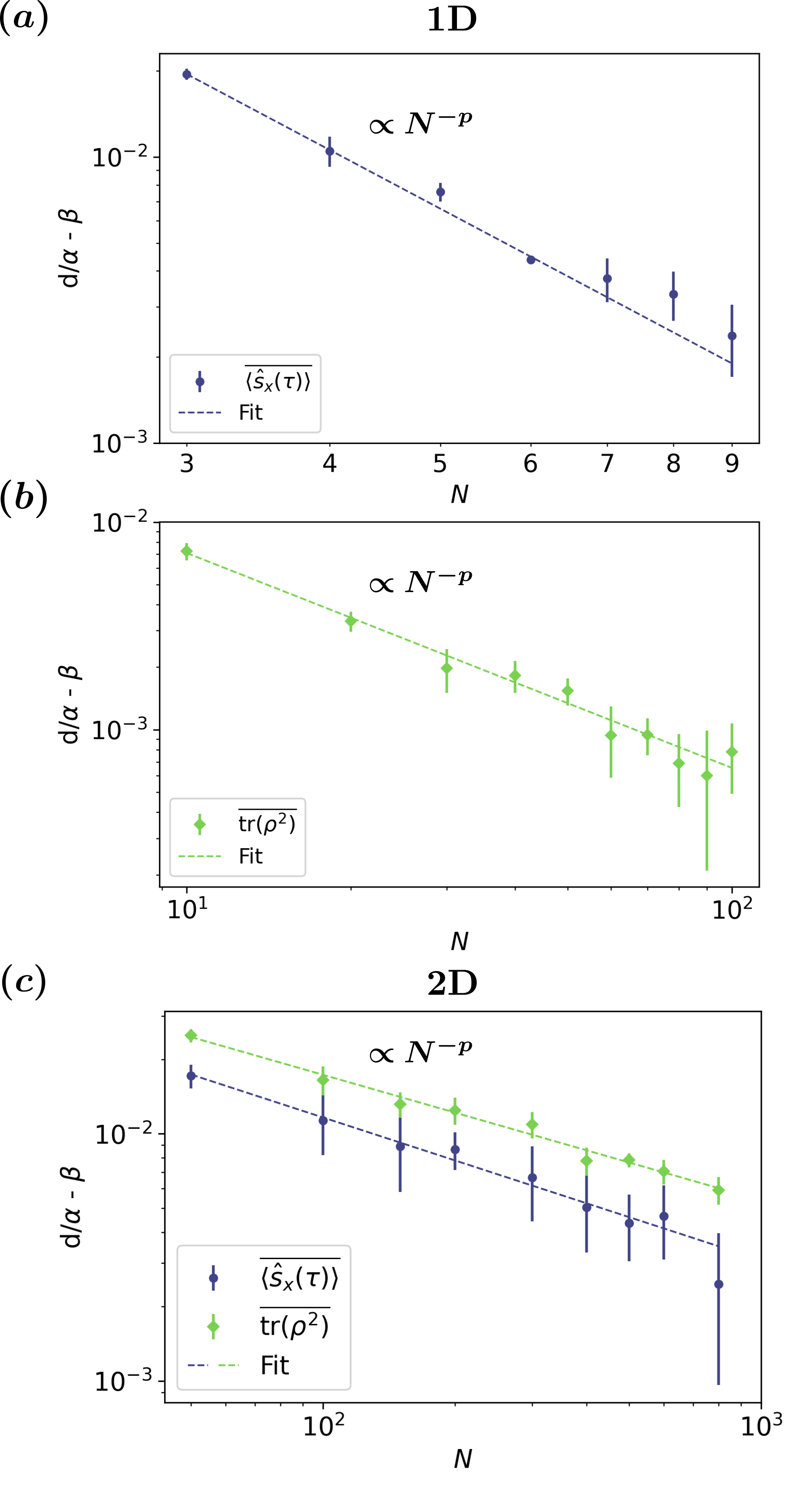}
    \caption{The scaling of the deviation from the thermodynamic limit solution $d/\alpha - \beta $ is shown for the case of $\alpha=6$ in  $d=1$ for ($\bm{a}$) the magnetization $\overline{\langle \hat{s}_x(\tau)}$ and ($\bm{b}$) purity $\overline{\mathrm{tr}(\rho^{2})}$. ($\bm{c}$) shows the relaxation for both quantities for $\alpha=6$ in $d=2$.} 
    \label{fig:supC}
\end{figure}

Similar to Fig.~2(b) of the main text we provide the scaling for $d=1$ and $d=2$ in Fig.~\ref{fig:supC}. The scaling is obtained down to an relative error of $\sim 1 \%$, where the uncertainty of the fit parameter starts to dominate remaining deviations, which is the reason for the different $N$ regimes in $d=1$ for $\overline{\langle \hat{s}_x(\tau)}$ and $\overline{\mathrm{tr}(\rho^{2})}$ respectively. All observed cases show power law convergence to the analytical result of the thermodynamic limit. 

Table~\ref{table1} shows the averaging parameters used in Fig.~2(c) of the main text. 

\begin{table}[h!]
    \centering
\begin{tabular}{|c|c|c|}
\hline 
    & \multicolumn{2}{|c|}{Magnetization/purity}  \\ \hline
    & $N$ & $N_s$ \\ \hline
    3D & 100 - 1300 & 200  \\
    2D & 50 - 800 & 200\\
    1D & 2-10 / 10-100 & 20000/4000\\ \hline
\end{tabular}
\label{table1}
\caption{Chosen parameter for Fig.~2(c) in the main text.}
\end{table}
 \clearpage

\end{document}